\documentclass[twocolumn,notitlepage,aps,longbibliography,pra]{revtex4-1}

\usepackage{amsmath}
\usepackage{amssymb}
\usepackage{bm}
\usepackage{bbm}
\usepackage{color}
\usepackage{graphicx}

\usepackage{hyperref}

\newcommand{\aoc}{{\tilde a}_0}

\definecolor{light}{gray}{0.90}
\definecolor{darker}{gray}{0.50}
\definecolor{dark}{gray}{0.30}

\newcommand{\dd}{{\rm d}}
\newcommand{\ee}{{\rm e}}
\newcommand{\ii}{{\rm i}}

\newcommand{\m}{{\rm m}}

\newcommand{\hate}{\hat{\mathrm{e}}}

\newcommand{\DD}{\mathrm{D}}
\newcommand{\NR}{\mathrm{NR}}
\newcommand{\TT}{\mathrm{T}}

\newcommand{\calN}{\mathcal{N}}
\newcommand{\calO}{\mathcal{O}}

\newcommand{\bbone}{\mathbbm{1}}

\begin{document}

\title{Algebraic Approach to Relativistic Landau Levels in the Symmetric Gauge}

\author{Ulrich D.~Jentschura}
%% \email{email: ulj@mst.edu}
\affiliation{Department of Physics and LAMOR,
Missouri University of Science and Technology,
Rolla, Missouri 65409, USA}

\begin{abstract}
We use an algebraic approach to the 
calculation of Landau levels for a 
uniform magnetic field in the symmetric 
gauge with a vector potential 
$\vec A = \tfrac12 (\vec B \times \vec r)$,
where $\vec B$ is assumed to be constant.
The magnetron quantum number
constitutes the degeneracy index.
An overall complex phase of the wave function,
given in terms of Laguerre polynomials,
is a consequence of the 
algebraic structure. The relativistic 
generalization of the treatment leads to 
fully relativistic bispinor Landau levels in the 
symmetric gauge, in a representation which 
writes the relativistic states 
in terms of their nonrelativistic limit,
and an algebraically accessible lower
bispinor component.
Negative-energy states and the massless limit are discussed.
The relativistic states can be used for a 
number of applications, including the calculation
of higher-order quantum electrodynamic binding corrections to 
the energies of quantum cyclotron levels.
\end{abstract}

\maketitle

\tableofcontents

%
% Introduction
%
\section{Introduction}
\label{sec1}

This paper is about nonrelativistic 
and relativistic Landau levels
for spin-$1/2$ particles,
pertinent to a uniform magnetic field.
We aim to develop algebraic relations, 
which describe the energetically 
degenerate Landau levels in terms
of cyclotron and magnetron quantum 
numbers. Furthermore, we aim to 
generalize the treatment to the 
fully relativistic domain, 
where the Landau levels become
eigenstates of the magnetically 
coupled Dirac equation.
Landau levels, including the relativistic
case, are important for many physical 
processes, first and foremost perhaps, 
in the context of the determination of 
fundamental physical constants,
for quantum cyclotron processes 
in Penning traps~\cite{BrGa1982,BrGa1986,Je2023limit,JeMo2023mag2}.
Among the many other application, we mention the 
description of the quantum Hall 
effect~\cite{Go2009,Go2011,Go2022,Ci2020,Ci2023},
2D quantum dots~\cite{BhMa2020},
particle production in the magnetars~\cite{MaEtAl2018},
and processes related to synchrotron 
radiation~\cite{SoTe1986,KadP2023}.

The constant, uniform magnetic field 
is assumed to be directed along the $z$ axis,
and used in the form
$\vec B_\TT = B_\TT \, \hate_z$,
where we use the subscript $\TT$ to denote
the possible application to a Penning trap~\cite{BrGa1982,BrGa1986}.
Landau levels are normally calculated~\cite{La1930}
in the Landau gauge $\vec A = -B y {\hat{\rm e}}_x$
[see Eq.~(112.1) of Ref.~\cite{LaLi1958vol3}].
Upon solving the corresponding 
Schr\"{o}dinger equation, one finds that
in the Landau gauge, energy eigenstates
are also eigenstates of 
the $x$ component $p_x$ of the momentum component.
The momentum component $p_x$
becomes a parameter which shifts the center of oscillation
for the effective harmonic oscillator 
potential, which acts onto the motion in the $y$ 
direction~[see Eq.~(112.5) of Ref.~\cite{LaLi1958vol3}].
In the Landau gauge, the value of $p_x$
does not affect the energy of the state
and becomes a continuous degeneracy index.

Here, we use the so-called symmetric 
gauge~\cite{PoCr1965,BhMa2020},
where the vector potential is taken as 
$\vec A = \tfrac12 (\vec B_\TT \times \vec r)$.
In the symmetric gauge, for the nonrelativistic case,
the eigenfunctions and 
energies are in principle known~\cite{PoCr1965,BhMa2020}.
Here, we augment the theory 
by identifying the raising and lowering operators
of the cyclotron and magnetron motions as
the determining dynamic variables of the problem.
This enables us to fully clarify 
the origin of the degeneracy of the 
(nonrelativistic) Landau levels in the symmetric gauge,
and greatly simplify the evaluation 
of matrix elements of eigenstates.

We also find the fully relativistic
generalization of the Landau levels,
corresponding to the relativistic quantum cyclotron
states in the limit of vanishing axial frequency.
The relativistic states can be used as input for higher-order
quantum electrodynamic calculations
of quantized-field effects in Penning traps~\cite{JeMo2023mag2},
and any physical systems where relativistic
effects become important.
Units with $\hbar = c = \epsilon_0 = 1$
are used throughout this paper.

%
% Nonrelativistic Problem
%
\section{Nonrelativistic Problem}
\label{sec2}

We start our treatment of Landau levels in the 
symmetric gauge, inspired 
by an electron in a Penning trap~\cite{BrGa1982,BrGa1986,WiMoJe2022},
where an additional electric quadrupole potential 
leads to axial confinement. 
In the limit of a vanishing axial 
frequency, the cyclotron and magnetron levels 
of the Penning trap become equal to Landau 
levels. In general, quantum cyclotron states are 
described~\cite{BrGa1982,BrGa1986,WiMoJe2022} by the quantum numbers $k \ell n s$,
where $k = 0,1,2,\dots$ is the axial excitation,
$\ell = 0,1,2, \dots$ is the magnetron quantum number,
$n= 0,1,2, \dots$ is the cyclotron quantum number,
and $s = \pm 1$ is the spin projection 
quantum number.

The kinetic momentum $\vec\pi_\TT  = \vec p - e \, \vec A$,
in the symmetric gauge,
can be written as
\begin{subequations}
\begin{align}
\vec \pi_\TT = & \;
\vec \pi_\parallel + 
\vec p_\perp \,,
\\[0.1133ex]
\vec \pi_\parallel = & \;
\vec p_\parallel - \frac{e}{2} \left( \vec B_\TT \times \vec r \right) \,,
\quad
\vec p_\parallel = p_x \, {\hat{\rm e}}_x +
p_y \, {\hat{\rm e}}_y \,,
\\[0.1133ex]
\vec B_\TT =& \; B_\TT  \, \hate_z \,.
\qquad
\vec p_\perp = p_z {\hat{\rm e}}_z \,.
\end{align}
\end{subequations}
Here, $e$ is the electron charge.
As already mentioned, we use the symmetric gauge
for the vector potential $\vec A = \tfrac12 (\vec B_\TT \times \vec r)$,
rather than
$\vec A' = -B y {\hat {\rm e}}_x$,
which yields the same magnetic field.
The gauge transformation $\Lambda = -\tfrac12 \, B_\TT x y$
leads from $\vec A = \tfrac12 (\vec B_\TT \times \vec r)$ 
to $\vec A'$.
We note that the physical interpretation of a 
wave function can depend on the gauge
(see p.~268 of Ref.~\cite{La1952}).
The choice $\vec A = \tfrac12 (\vec B_\TT \times \vec r)$
leads to wave functions which are confined in the $x$ and $y$
directions, while, with the choice $\vec A'$,
one obtains unconfined wave functions in the $y$ direction.

Furthermore, the choice $\vec A = \tfrac12 (\vec B_\TT \times \vec r)$
is generalizable to the case of
a nonvanishing additional electric quadrupole trap field~\cite{BrGa1982,BrGa1986}.
The nonrelativistic Hamiltonian is given as follows,
\begin{align}
H_0 =& \; 
\frac{(\vec\sigma \cdot \vec\pi_\TT)^2}{2m}
= H_\parallel + \frac{\omega_c}{2} \sigma_z 
+ \frac{p_z^2}{2m} \,,
\\[0.1133ex]
\omega_c =& \; - \frac{e \, B_\TT}{m} 
= \frac{|e| \, B_\TT}{m} \,.
\end{align}
Here, $|e| = -e$ is the modulus of the electron charge.
The spin-independent 
part of the Hamiltonian relevant for the $xy$ plane is
\begin{subequations}
\begin{align}
\label{defHxy}
H_\parallel =& \; 
\frac{\vec \pi_\parallel^{\,2}}{2 m} =
\frac{\vec p_\parallel^{\,2}}{2 m} +
\frac{\omega_c}{2} L_z +
\frac{m \omega_c^2}{8} \, \rho^2
\nonumber\\[0.1133ex]
=& \; \omega_c \left( a_c^\dagger \, a_c + \frac12 \right) \,,
\end{align}
\end{subequations}
where the $a_c^\dagger$ and $a_c$
are the raising and lowering operators
of the cyclotron motion.
We have set the electric quadrupole potential
of the trap equal to zero, 
wherefore, in the sense of Ref.~\cite{BrGa1982,BrGa1986},
the magnetron frequency vanishes.
The cyclotron ($c$) and magnetron ($\m$) raising 
($a_c^\dagger$ and ($a_\m^\dagger$)
and lowering operators ($a_c$ and $a_\m$)
are given as follows,
\begin{subequations}
\label{landau}
\begin{align}
a_c =& \; \frac{1}{\sqrt{2}} \,
\left( \aoc \, p_x - \ii \aoc \, p_y -
\frac{\ii \, x}{2 \, \aoc} -
\frac{y}{2 \, \aoc} \right) \,,
\\[0.1133ex]
a^\dagger_c =& \; \frac{1}{\sqrt{2}} \,
\left( \aoc \, p_x + \ii \aoc \, p_y +
\frac{\ii \, x}{2 \, \aoc} -
\frac{y}{2 \, \aoc} \right) \,,
\\[0.1133ex]
a_\m =& \; \frac{1}{\sqrt{2}} \,
\left( \aoc \, p_x + \ii \aoc \, p_y -
\frac{\ii \, x}{2 \, \aoc} +
\frac{y}{2 \, \aoc} \right) \,,
\\[0.1133ex]
a_\m^\dagger =& \; \frac{1}{\sqrt{2}} \,
\left( \aoc \, p_x - \ii \aoc \, p_y +
\frac{\ii \, x}{2 \, \aoc} +
\frac{y}{2 \, \aoc} \right) \,.
\end{align}
\end{subequations}
In these formulas, the generalized (magnetic) Bohr radius is
(we temporarily restore factors of $\hbar$ and $c$)
\begin{equation}
\aoc = \sqrt{\frac{m c^2}{\hbar \omega_c}} \,
\frac{\hbar}{ m c}
= \frac{\hbar}{\alpha_c m c} 
= \sqrt{ \frac{\hbar}{|e| B_\TT}} = \ell_B \,,
\end{equation}
where we have set $\alpha_c = \sqrt{\hbar \omega_c/(m c^2)}$
($\alpha_c$ is a generalized fine-structure constant),
and $\ell_B$ is the magnetic length~\cite{Go2009}.
The relations~\eqref{landau} can be inverted,
\begin{subequations}
\label{landau2}
\begin{align}
p_x =& \; \frac{1}{2 \sqrt{2} \aoc} \,
\left[ a_c + a_c^\dagger + a_\m + a_\m^\dagger \right] \,,
\\
p_y =& \; \frac{\ii}{2 \sqrt{2} \aoc} \,
\left[ a_c - a_c^\dagger - a_\m + a_\m^\dagger \right] \,,
\\
x =& \; \frac{\ii \aoc}{\sqrt{2}} \,
\left[ a_c - a_c^\dagger + a_\m - a_\m^\dagger \right] \,,
\\
y =& \; -\frac{\aoc}{\sqrt{2}} \,
\left[ a_c + a_c^\dagger - a_\m - a_\m^\dagger \right] \,.
\end{align}
\end{subequations}
The nonvanishing commutators are as follows,
\begin{equation}
[ a_c, a^\dagger_c ] =
[ a_\m, a^\dagger_\m ] = 1 \,.
\end{equation}
All other commutators vanish,
which means that, in particular,
the cyclotron and magnetron 
excitation numbers, together with the $z$ component
of the momentum and the spin projection quantum number 
$s$, form a complete set of commuting 
observables for the nonrelativistic spin-independent
problem. The eigenfunctions are given as follows,
\begin{subequations}
\label{Eklns}
\begin{align}
H_\NR \, \psi_{k n \ell s}(\vec r) =& \;
E_\NR \, \psi_{k n \ell s}(\vec r)  \,,
\\[0.1133ex]
E_\NR =& \; \omega_c \, \left( n + \frac{s + 1}{2} \right) 
+ \frac{k^2}{2 m} \,.
\end{align}
\end{subequations}
Here, $k \in \mathbbm{R}$ is a continuous quantum number 
characterizing the momentum component in the $z$ direction.
The energy eigenvalue is independent of the 
magnetron quantum number $\ell$.
The momentum in the $z$ component and the 
spin component of the eigenstate can be split off,
\begin{subequations}
\label{waveklns}
\begin{align}
\psi_{k \ell n s}(\vec r) =& \;
\psi_{n \ell}(\vec\rho)  \, 
\frac{\ee^{\ii k z}}{\sqrt{2\pi}} \, \chi_s \,,
\\[0.1133ex]
\vec\rho =& \; x \, \hate_x + y \, \hate_y \,,
\\[0.1133ex]
\chi_{+1} =& \; \left( 
\begin{array}{c} 1 \\ 0 
\end{array} \right) \,,
\qquad
\chi_{-1} = \left( 
\begin{array}{c} 0 \\ 1 
\end{array} \right) \,.
\end{align}
\end{subequations}

According to Refs.~\cite{BrGa1982,BrGa1986,WiMoJe2022},
one uses the quantum numbers $k \ell n s$ 
(in this, alphabetically inspired, sequence) by convention.
However, for the spin-independent,
two-dimensional, nonrelativistic case,
we use the designation $n \ell$, in correspondence with the 
hydrogen atom 
where the first index indicates the quantum
number which determines the energy
(for a comprehensive discussion,
see Chap.~4 of Ref.~\cite{JeAd2022book}).
One may search for the general formula for the nonrelativistic 
eigenfunction $\psi_{n \ell}(\vec\rho)$,
which fulfills the equation
\begin{equation}
\label{Enl}
H_\parallel \, \psi_{n \ell}(\vec\rho) =
E_n \, \psi_{n \ell}(\vec\rho) \,,
\qquad
E_n = \omega_c \left( n + \frac12 \right) \,,
\end{equation}
where $H_\parallel$ is given in Eq.~\eqref{defHxy}.
The solution reads as follows,
\begin{align}
\label{wavenl}
\psi_{n \ell}(\vec\rho) =& \;
\frac{ 2^{-\tfrac12 (|n-\ell|+1)} }{ \sqrt{\pi} \, \aoc } \,
\sqrt{ \frac{ \min(n, \ell)! }{ \max(n, \ell)! }} \,
\left( \frac{\rho}{\aoc} \right)^{|n - \ell|} 
\nonumber\\
& \; 
\times 
\ii^{ | n - \ell | } \,
L^{| n - \ell|}_{\min(n, \ell)}%
\left( \frac12 \, \left( \frac{\rho}{\aoc} \right)^2  \right) \,
\nonumber\\
& \; \times \ee^{\ii (n-\ell) \varphi} \,
\exp\left[ - \frac14 \, \left( \frac{\rho}{\aoc} \right)^2  \right] \,.
\end{align}
Here, $\rho = | \vec \rho |$ and $\varphi = \arctan(y/x)$.
The complex phase $\ii^{ | n - \ell | }$ is a 
consequence of the application of the 
raising operators on the ground states 
[see also Eq.~\eqref{raise_ground}].

A comparison of Eq.~\eqref{wavenl} with the literature 
is indicated.
In Eq.~(2) of Ref.~\cite{BaCNPa1976},
one obtains a corresponding
result. The spin-independent wave function obtained
in Ref.~\cite{BaCNPa1976}
is equal to our wave 
function if one replaces 
$k$ (in the notation of Ref.~\cite{BaCNPa1976}
by $\min(n, \ell)$, and 
and $m$ (in the notation of Ref.~\cite{BaCNPa1976})
by $n - \ell$ (in our notation).
Furthermore, one may point out that 
the wave functions given in
Eq.~(2) of Ref.~\cite{BaCNPa1976} are specialized to the 
case $\omega_c = 2$.
In Ref.~\cite{Ci2020}, this restriction is 
lifted. The parameter $l_0$ defined in 
Eq.~(28) of Ref.~\cite{Ci2020} is equal to 
our parameter ${\tilde a}_0$,

In Eq. (A31) of Ref.~\cite{BhMa2020},
the authors obtain a corresponding
result.  Their parameter 
$\ell$ (which, in the conventions
of Ref.~\cite{BhMa2020} can become negative) is 
equal to our $n-\ell$.
The inclusion of additional electric fields into the
formalism is discussed in 
Refs.~\cite{BrGa1982,BrGa1986,WiMoJe2022} 
and in Ref.~\cite{EdAu2019}.

\begin{figure}[t!]
\begin{center}
\begin{minipage}{1.0\linewidth}
\begin{center}
\includegraphics[width=0.87\linewidth]{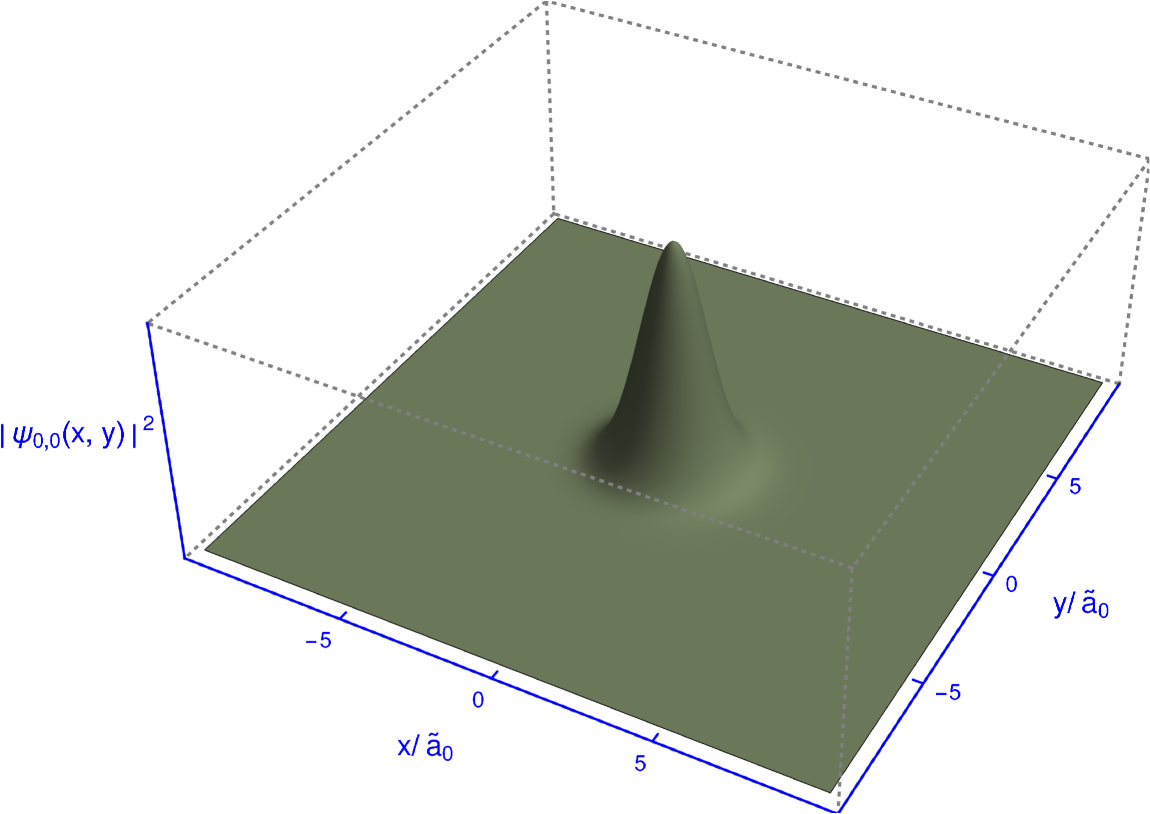}
\caption{\label{fig1}
We display the probability density 
$| \psi_{(n=0),(\ell=0)}(\vec \rho) |^2$
of the cyclotron ground state,
as given in Eq.~\eqref{wavenl}.
It is radially symmetric and nonvanishing at
the origin.}
\end{center}
\end{minipage}
\end{center}
\end{figure}

\begin{figure}[t!]
\begin{center}
\begin{minipage}{1.0\linewidth}
\begin{center}
\includegraphics[width=0.82\linewidth]{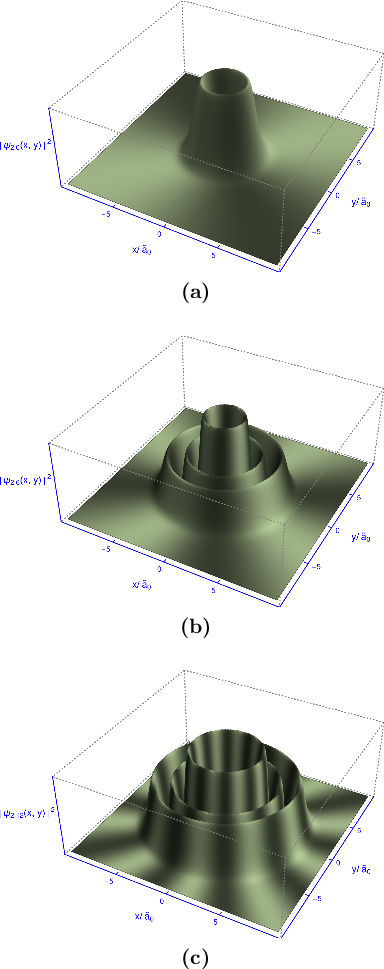}
\caption{\label{fig2}
The cyclotron states with $n=2$ all have 
the same nonrelativistic energy $\tfrac52 \omega_c$.
The probability density $| \psi_{(n=2) \ell}(\vec \rho)|^2$ 
vanishes at the origin.
With increasing magnetron excitation $\ell = 0, 6, 12$,
the states spread away from the origin.
The azimuthal dependence of the 
wave function is given by the factor $\exp[\ii(n-\ell) \, \varphi]$.
The plot surfaces are shaded according to 
the imaginary part of the wave function,
with lighter regions indicating a positive imaginary part,
and darker regions indicating a negative imaginary part
of the wave function.}
\end{center}
\end{minipage}
\end{center}
\end{figure}

Our formalism clearly identifies
the cyclotron excitation, which is linked to the quantum
number $n$ that determines the energy,
and the magnetron quantum number $\ell$,
which does not shift the energy but shifts the  
orbit further outward with increasing $\ell$  
for given $n$ (see Figs.~\ref{fig1} and~\ref{fig2}). 
When one increases $\ell$  
by one, the wave function acquires 
a phase factor $\exp(-\ii \varphi)$.
The cyclotron and magnetron excitations can be
associated with the action of mutually commuting  
raising operators, acting on the ground state,
and the quantum numbers $n$ and $\ell$ belong to mutually 
commuting observables.

Indeed, the nonrelativistic
eigenfunctions can be obtained from the 
ground state by the operation of the 
raising operators,
\begin{subequations}
\begin{align}
\label{raise_ground}
\psi_{ n \ell }(\vec \rho) =& \; 
\frac{1}{\sqrt{ n! \, \ell! }} \,
( a_c^\dagger )^n \,
( a_\m^\dagger )^\ell \, 
\psi_{  0 0  }(\vec \rho)  \,,
\\
\psi_{  0 0  }(\vec \rho)  =& \;
\frac{1}{\aoc \, \sqrt{2 \pi}} \, 
\exp\left( - \frac{\rho^2}{4 \aoc^2} \right)  \,.
\end{align}
\end{subequations}
The number operators count the number of 
cyclotron and magnetron excitations,
\begin{equation}
\label{raise}
a_c^\dagger \, a_c \, | n, \ell \rangle = 
n \, | n, \ell \rangle \,,
\qquad
a_\m^\dagger \, a_\m \, | n, \ell \rangle = 
\ell \, | n, \ell \rangle  \,.
\end{equation}
The wave functions $\psi_{ n \ell }(\vec \rho)$ are 
eigenfunctions of the Hamiltonian $H_\parallel$, 
while their energy eigenvalue only depends on 
the cyclotron quantum number $n$,
and is independent of the magnetron quantum number $\ell$.
The magnetron quantum number $\ell = 0,1,2,\dots$ acts
as a degeneracy index for the Landau level.
Here, the magnetron degeneracy index is countable
and leads to a more intuitive form of the wave
function, which is normalizable to unity.
Wave functions are orthonormal,
\begin{equation}
\label{ortovox1}
\int \dd^2 \rho \, \psi^*_{n \ell}(\vec \rho) \, 
\psi_{n' \ell'}(\vec \rho) = 
\delta_{n n'} \, \delta_{\ell \ell'} \,.
\end{equation}
A total of four example wave functions are 
plotted in Figs.~\ref{fig1} and~\ref{fig2}.
Furthermore, one has the relation
\begin{equation}
\label{ortovox2}
\int \dd^3 r \, \psi^\dagger_{k \ell n s}(\vec \rho) \, 
\psi_{k' \ell' n' s'}(\vec \rho) = 
\delta( k - k') \, \delta_{n n'} \, 
\delta_{\ell \ell'} \,  \delta_{s s'} \,.
\end{equation}
(For the spinor case, we need to use $\psi^\dagger$ because 
it is a two-component wave function.)
With increasing magnetron quantum number, 
the states spread further away from the
origin, as is evident from the 
expectation value
\begin{equation}
\int \dd^2 \rho \, \rho^2 \, | \psi_{n \ell}(\vec \rho)|^2 
= (2 + 2 n + 2 \ell) \aoc^2 \,.
\end{equation}
This trend is depicted in Fig.~\ref{fig2}.

%
% Relativistic Landau Levels
%
\section{Relativistic Landau Levels}
\label{sec3}

Before we indulge in the calculations, let us 
remember that the problem of a relativistic 
electron coupled to a uniform electric or magnetic
field, has been discussed in the literature before,
notably, in a comprehensive treatise given in 
Ref.~\cite{De1953}. 
In unnumbered equations between
Eqs.~(10) and (11)
of Sec.~3 of Chap.~3 of Ref.~\cite{De1953}, 
the relativistic eigenstate
of the magnetically coupled Dirac equation is 
given in terms of parabolic cylinder functions
(Weber functions, denoted as $D_n$).
The vector potential is taken in the Landau gauge.
Alternative discussions of the 
relativistic states are given in 
Sec.~4 of Chap.~26 of Ref.~\cite{SoTe1986}
and in Sec.~6 of Chap.~2 of Ref.~\cite{BaGi1990}.
Our goal here is to find a representation of the 
relativistic eigenstates which allows us to 
clearly identify the connections of the 
nonrelativistic limit with the fully relativistic
state.

We here aim to find the relativistic Landau levels
in the symmetric gauge.
Let us recall the nonrelativistic (NR) Hamiltonian
and its eigenstates,
\begin{align}
H_\NR \psi_\NR =& \; E_\NR \psi_\NR \,,
\quad
H_\NR = \frac{( \vec\sigma \cdot \vec \pi_\TT)^2}{2m} \,,
\\[0.1133ex]
\qquad
E_\NR =& \; \omega_c \left( n + \frac{s + 1}{2} \right)
+ \frac{k^2}{2 m} \,.
\end{align}
The kinetic momentum in the trap is
$\vec \pi_\TT = \vec p - e \vec A_\TT$,
while the vector potential $\vec A_\TT$ for the 
uniform magnetic field of the trap can be written as
\begin{equation}
\vec \pi_\TT = \vec p - e \vec A_\TT \,,
\qquad
\vec A_\TT = \tfrac12 (\vec B_\TT \times \vec r) \,.
\end{equation}
The Dirac Hamiltonian for a uniform magnetic field
is 
\begin{align} 
H_\DD =& \; \vec\alpha \cdot \vec \pi_\TT + \beta \, m \,,
\\[0.1133ex]
\vec \alpha =& \; 
\left( \begin{array}{cc} 
0 & \vec \sigma \\
\vec \sigma & 0 
\end{array} \right) \,,
\quad
\beta = 
\left( \begin{array}{cc} 
\bbone_{2 \times 2} & 0 \\
0 & -\bbone_{2 \times 2} \\
\end{array} \right) \,,
\end{align} 
where we use the Dirac matrix in the 
Dirac representation.
We search for bispinor solutions of the form
\begin{subequations} 
\begin{equation} 
\label{Wave1}
H_\DD \, \Psi = E_\DD \, \Psi \,,
\qquad
\Psi = 
\left( \begin{array}{c} 
\psi_1 \\
\psi_2 
\end{array} \right) \equiv
\Psi_{k \ell n s} \,.
\end{equation} 
We note that $\Psi$ denotes the relativistic bispinor
wave function, while $\psi$ is the nonrelativistic 
counterpart. We use the following 
{\em ansatz} for the relativistic eigenstates,
\begin{equation} 
\label{Wave2}
\psi_1 = \calN \, \psi_\NR \,,
\qquad
\psi_2 = \calN \, \frac{\vec\sigma \cdot \vec \pi_\TT}{E_D + m} \psi_\NR \,.
\end{equation} 
\end{subequations} 
Acting with $H_\DD$ on $\Psi$, one obtains
\begin{align} 
\label{HDPsi}
H_\DD \Psi = & \;
%
%%% \left( \begin{array}{cc} 
%%% m \bbone_{2 \times 2} & \vec \sigma \cdot \vec \pi_\TT \\
%%% \vec \sigma \cdot \vec \pi_\TT & -m \bbone_{2 \times 2} \\
%%% \end{array} \right) 
%%% \cdot
%%% \left( \begin{array}{c} 
%%% \psi_1 \\ \psi_2 
%%% \end{array} \right) 
%%% \nonumber\\[0.1133ex]
%
\calN \, 
\left( \begin{array}{cc}
m \bbone_{2 \times 2} & \vec \sigma \cdot \vec \pi_\TT \\
\vec \sigma \cdot \vec \pi_\TT & -m \bbone_{2 \times 2} \\
\end{array} \right)
\left( \begin{array}{c} \psi_\NR \\ 
\frac{\vec\sigma \cdot \vec \pi_\TT}{E_\DD + m} \psi_\NR 
\end{array} \right)
\nonumber\\[0.1133ex]
=& \; \calN \, \left( \begin{array}{cc} 
m \left( 1 + \dfrac{2 E_\NR }{E_\DD + m}  \right)
\psi_\NR \\[1.1133ex]
\vec \sigma \cdot \vec \pi_\TT 
\left( 1 - \dfrac{m}{E_\DD + m} \right) \psi_\NR
\end{array} \right) 
\nonumber\\[0.1133ex]
\mathop{=}^{\mbox{!}} & \;
\calN \, E_\DD \, \left( \begin{array}{c} \psi_\NR \\[1.1133ex]
\dfrac{\vec\sigma \cdot \vec \pi_\TT}{E_\DD + m} \psi_\NR 
\end{array} \right) = E_D \, \Psi \,.
\end{align}
From the secular equation for the upper component, 
one obtains
\begin{equation} 
E_\DD = m \left( 1 + \frac{2 E_\NR }{E_\DD + m}  \right)
= m + \frac{2 m E_\NR }{E_\DD + m} \,.
\end{equation} 
This can be rewritten as follows,
\begin{equation} 
E_\DD - m = \frac{2 m E_\NR }{E_\DD + m}  \,,
\qquad
E_\DD^2 - m^2 = 2 m E_\NR \,.
\end{equation} 
The Dirac energy $E_\DD$ is obtained
as follows,
\begin{align}
\label{EDD}
E_\DD =& \; \sqrt{ m^2 + 2 m E_\NR }
= m \sqrt{ 1 + \frac{ 2 E_\NR }{m} }
\nonumber\\[0.1133ex]
=& \; m \sqrt{ 1 + \frac{ \omega_c }{m} \left( 2 n + s + 1 \right)
+ \frac{k^2}{m^2} } \,.
\end{align}
Somewhat surprisingly, the state with 
quantum numbers $n=0$ and $s=-1$, and $k=0$,  has exact 
rest mass energy even in the fully relativistic formalism.
A check of the eigenvector 
property for the lower component is successful,
We start from the
expression in round brackets (lower component)
in the second line of Eq.~\eqref{HDPsi} and obtain:
\begin{equation} 
\left( 1 - \frac{m}{E_\DD + m} \right) =
\frac{E_\DD}{E_\DD + m} =
E_\DD \, \frac{1}{E_\DD + m} \,.
\end{equation} 
In view of the Taylor expansion
$\sqrt{1 + \epsilon} = 
1 + \frac{\epsilon}{2} 
- \frac{\epsilon^2}{8} 
+ \frac{\epsilon^3}{16}  + \calO(\epsilon^4)$,
we can write
for positive energy in the limit $k \to 0$,
\begin{multline} 
E_\DD(k \to 0) = m + 
\frac{ \omega_c }{2 m} \left( 2 n + s + 1 \right) 
\\[0.1133ex]
- \frac18 
\left( \frac{ \omega_c }{m} \left( 2 n + s + 1 \right) \right)^2 
+ \frac{1}{16}
\left( \frac{ \omega_c }{m} \left( 2 n + s + 1 \right) \right)^3 \,.
\end{multline} 
This is consistent with Eqs. (90), (96) and (105) 
of Ref.~\cite{WiMoJe2022}.
We set $\omega_c = \alpha_c^2 m$,
where $\alpha_c$ is the cyclotron coupling constant.
The normalization of the relativistic states
according to :
$\int \dd^3 r \, | \psi_\NR |^2 = \int \dd^3 r | \Psi |^2 = 1$
leads to the relation
\begin{align} 
\label{Wave3}
\int \dd^3 r \; \Psi^\dagger(\vec r) \, \Psi(\vec r) =& \;
\calN^2 \left[ 1 + \frac{2 m E_\NR}{(E_\DD + m)^2} \right] 
\mathop{=}^{\mbox{!}} 1 \,,
\\[0.1133ex]
\label{defcalN}
\calN =& \; \left[ 1 + \frac{2 m E_\NR}{(E_\DD + m)^2} \right]^{-1/2} \,.
\end{align} 
In the above formulas for the relativistic 
states, the dependence on the quantum numbers $k\ell n s$
has been suppressed for notational simplicity.
Restoring the quantum numbers, one finds that 
\begin{equation}
\label{ortovox}
\int \dd^3 r \, \Psi^\dagger_{k \ell n s}(\vec \rho) \, 
\Psi_{k' \ell' n' s'}(\vec \rho) = 
\delta( k - k') \, \delta_{n n'} \, 
\delta_{\ell \ell'} \,  \delta_{s s'} \,.
\end{equation}
%%% 
%%% Expand[ m^2 ( 1 + (2 ENR)/(EE + m) )^2 + 2 m ENR ( 1 - m/(EE + m) )^2 ]
%%% xx1 =  m^2 ( 1 + (2 ENR)/(EE + m) )^2 + 2 m ENR ( 1 - m/(EE + m) )^2 
%%% xx2 =  m ( m + 2 ENR) + 2 ENR m^2 (m + 2 ENR)/ (EE + m)^2
%%% Together[xx1-xx2]
%%% 
For completeness, it is perhaps useful to 
remark that, if desired, one can easily 
write the entire operator $\vec \sigma \cdot \vec \pi_\TT$
in terms of cyclotron, magnetron and spin ladder operators,
and express the lower component $\psi_2$ of the 
fully relativistic state, in terms of 
nonrelativistic eigenstates of raised and lowered 
quantum numbers. 

%
% Negative--Energy States
%
\section{Negative--Energy States}
\label{sec4}

In Dirac theory, there is an energy gap 
between positive-energy states with $E \geq m$
and negative-energy states with $E \leq -m$.
It is relatively straightforward to check that the states
[cf.~Eqs.~\eqref{Wave1} and~\eqref{Wave2}]
\begin{equation}
\Phi = \calN \left( \begin{array}{c}
-\dfrac{\vec\sigma \cdot \vec \pi_\TT}{E_\DD + m} \, \psi_\NR \\[3ex]
\psi_\NR
\end{array} \right)
\end{equation}
with the same normalization factor $\calN$ as
for positive-energy states [see Eq.~\eqref{defcalN}],
has the property
\begin{equation}
H_\DD \, \Phi = - E_\DD \Phi
\end{equation}
and thus constitutes the negative-energy 
(antiparticle) state of Dirac theory.
As compared to the positive-energy 
state, the upper and lower components of the 
Dirac bispinor are interchanged, 
and the upper spinor receives a 
minus sign.

%
% Massless Limit
%
\section{Massless Limit}
\label{sec5}

In the limit $m \to 0$, the gap between positive-energy 
and negative-energy states vanishes. 
We recall that
\begin{equation}
\omega_c m = |e| B_\TT = \ell_B^{-2} \,.
\end{equation}
The Dirac energy~\eqref{EDD} attains the 
massless limit
\begin{align}
E_\DD \mathop{=}^{m \to 0}
\sqrt{ \ell_B^{-2}  \left( 2 n + s + 1 \right) + k^2 } \,.
\end{align}
The square-root dependence on the 
quantum numbers confirms the 
corresponding dependence, obtained 
on the basis of the Weyl equation,
given in Eq.~(2.23) 
of Ref.~\cite{Go2009}.  
The appropriate zero-mass limit of the 
normalization factor is 
\begin{equation}
\calN \mathop{=}^{m \to 0}  \frac{1}{\sqrt{2}} \,,
\end{equation}
and the positive-energy solution is
\begin{equation}
\Psi \mathop{=}^{m \to 0}
\left( \begin{array}{c}
\dfrac{1}{\sqrt{2}} \, \psi_\NR  
\\[3.3ex]
\dfrac{1}{\sqrt{2}} \, \dfrac{\vec\sigma \cdot \vec \pi_\TT}{E_D} 
\psi_\NR 
\end{array} \right) \,.
\end{equation}
The negative-energy solution is 
\begin{equation}
\Phi \mathop{=}^{m \to 0}
\left( \begin{array}{c}
-\dfrac{1}{\sqrt{2}} \, \dfrac{\vec\sigma \cdot \vec \pi_\TT}{E_D}
\psi_\NR
\\[3.3ex]
\dfrac{1}{\sqrt{2}} \, \psi_\NR
\end{array} \right) \,.
\end{equation}

%
% Conclusions
%
\section{Conclusions}
\label{sec6}

We have considered the calculation of 
both nonrelativistic as well as fully relativistic 
Landau levels in the symmetric gauge where
$\vec A = \tfrac12 (\vec B_\TT \times \vec r)$,
based on an algebraic approach.
The spin-independent part of the 
nonrelativistic states finds a natural form in terms
of the cyclotron quantum number $n$, and
of the magnetron quantum number $\ell$.
We find a universal representation of the 
spinless nonrelativistic quantum state wave function
with quantum numbers
$n$ (cyclotron) and
$\ell$ (magnetron) in Eq.~\eqref{wavenl}.
After including the spin and the axial motion,
we find the nonrelativistic spinor state with 
quantum numbers $n$ and $\ell$,
and $k$ (axial) and $s$ (spin),
as given in Eq.~\eqref{waveklns}.
For the nonrelativistic wave
function $\psi = \psi_{k\ell n s}$,
the spin-independent form is given 
in Eq.~\eqref{wavenl}, and
the nonrelativistic spinor form is given 
in Eq.~\eqref{waveklns}.
The orthonormality relations are given in 
Eqs.~\eqref{ortovox1} (for the 
spin-independent part) and
in Eq.~\eqref{ortovox2} (for the spinor
wave function).
While the magnetron quanta are raised from
the ground state by the magnetron 
raising operator $a_\m^\dagger$ 
[see Eq.~\eqref{raise}],
the magnetron quantum number $\ell$ 
does not enter the formula for the 
energy of the Landau level [see
Eqs.~\eqref{Enl} and~\eqref{Eklns}].
Inverting the relations given in 
Eq.~\eqref{landau}, one obtains a 
representation of the $x$, $y$, $p_x$ 
and $p_y$ operators in terms of the 
raising and lowering
operators of the cyclotron and magnetron motions
[see Eq.~\eqref{landau2}].

In the context of the quantum Hall effect,
Landau levels have been considered 
in both the Landau as well as the 
symmetric gauges~\cite{Go2009}.
They are also relevant to 
quantum cyclotron states in Penning traps~\cite{BrGa1982,BrGa1986}.
Indeed, in the limit of vanishing
axial confinement, the quantum cyclotron states
approach the Landau levels in the symmetric 
gauge and our formulas are essential elements
in the discussion of higher-order quantum
electrodynamic corrections to 
quantum cyclotron energy levels~\cite{JeMo2023mag2}.

Based on an essential generalization
of the nonrelativistic problem (see Sec.~\ref{sec2} and Ref.~\cite{Go2009}),
we calculate the fully relativistic Landau levels
in the symmetric gauge, in Sec.~\ref{sec3}.
These relativistic states constitute solutions of the 
magnetically coupled relativistic Dirac equation.
The same quantum numbers $k$, $\ell$, $n$ and $s$,
that we found for the nonrelativistic problem,
characterize the relativistic state.
The relativistic wave function is given in
Eqs.~\eqref{Wave1} and~\eqref{Wave2},
with the normalization given in Eq.~\eqref{Wave3}.
The fully relativistic form is needed
as an essential ingredient in the calculation
of quantum electrodynamic corrections to 
quantum cyclotron energy levels~\cite{JeMo2023mag2}.
We mention relativistic quantum dynamics
in synchrotrons as a further potential area
of application.
Negative-energy states are discussed in Sec.~\ref{sec4},
and the massless limit of the Dirac bispinor solutions 
is discussed in Sec.~\ref{sec5}.

\section*{Acknowledgments}

The author acknowledge insightful
conversations with Istv\'{a} N\'{a}ndori,
and Professor Gerald Gabrielse.
Support from the National Science Foundation
(grant PHY--2110294) is gratefully acknowledged.
Furthermore, support from the
Templeton Foundation (Fundamental Physics Black Grant,
Subaward 60049570 of Grant ID \#{}61039),
is also gratefully acknowledged.

\end{document}